\documentclass[aps,prb,twocolumn,superscriptaddress,english]{revtex4-2}
\usepackage[utf8]{inputenc}

\usepackage{amsmath}
\usepackage{amssymb}
\usepackage{graphicx}
\usepackage{xcolor}
\usepackage{siunitx}
\usepackage{lipsum}
\usepackage{lineno}
\usepackage{tabularx}
\usepackage{orcidlink}
\usepackage[version=4]{mhchem}
\usepackage{chemformula}
\usepackage{isotope}

\usepackage{amssymb}
\usepackage{amsmath}
\usepackage{mathtools}
\usepackage{physics}

\usepackage{hyperref}

\makeatletter
\def\@fnsymbol#1{\ensuremath{\ifcase#1\or \dagger\or *\or \ddagger\or
		\mathsection\or \mathparagraph\or \|\or **\or \dagger\dagger
		\or \ddagger\ddagger \else\@ctrerr\fi}}
\makeatother

\begin{document}
	
	\author{Caja Annweiler \orcidlink{0009-0004-7201-859X}}\email{caja.annweiler@physik.hu-berlin.de}
	\affiliation{Heidelberg University, Institute for Theoretical Physics, Philosophenweg 19, 69120 Heidelberg, Germany}
	\author{Simone Di Cataldo \orcidlink{0000-0002-8902-0125}} \email{simonde.dicataldo@uniroma1.it}
	\affiliation{Dipartimento di Fisica, Sapienza Universit\`a di Roma, 00185 Roma, Italy}
	\author{Maurits W. Haverkort \orcidlink{0000-0002-7216-3146}} \email{M.W.Haverkort@thphys.uni-heidelberg.de}
	\affiliation{Heidelberg University, Institute for Theoretical Physics, Philosophenweg 19, 69120 Heidelberg, Germany}
	\author{Lilia Boeri \orcidlink{0000-0003-1186-2207}} \email{lilia.boeri@uniroma1.it}
	\affiliation{Dipartimento di Fisica, Sapienza Universit\`a di Roma, 00185 Roma, Italy}

	\title{A method for the automatic generation of a minimal basis set of structural templates for material phase-space exploration}
	
	%QUICKSILVER: Quantum Simulation-based Identification of Key Structural Templates for Binary Material Exploration
	%STARGATE: Structural Template Analysis and Generation via Quantum Exploration
	%QUARKS: Quantum Analysis for Research and Knowledge of Structural Templates
	%GENESIS: In-silico Generation of Essential Structural Templates for Material Space Exploration
	%STRUCTGEN: Structural Template Generation  for in-silico Material Space Exploration
	%MATERIALIZE: Method for In-silico Generation of Minimal Basis of Structural Templates for Material Space Exploration
	
	\date{\today}
	\begin{abstract}
		We present a novel method for predicting binary phase diagrams through the automatic construction of a minimal basis set of representative templates. The core assumption is that any materials space can be divided into a small number of regions with similar chemical tendencies and bonding properties, and that a minimal set of templates can efficiently represent the key chemical trends across the different regions. By combining data-driven techniques with ab-initio crystal structure prediction, we can efficiently partition the materials space and construct templates reflecting variations in chemical behavior. Preliminary results demonstrate that our method predicts binary convex hulls with accuracy comparable to resource-intensive EA searches, while achieving a significant reduction in computational time (by a factor of 25). The method can be extended to ternary and multinary systems, enabling efficient high-throughput exploration and mapping of complex material spaces. By providing a transformative solution for high-throughput materials discovery, our approach paves the way for uncovering advanced quantum materials and accelerating in silico design.
	\end{abstract}

	\maketitle

	\section*{Introduction} 
	Material research is undergoing a transformative phase.
	Advances in non-equilibrium and high-pressure synthesis techniques have dramatically broadened the scope of materials science to levels 
	thought unattainable just a decade ago.
	Moreover,  rapid developments in first-principles simulations,
	offer the unprecedented possibility to explore in silico 
	vast regions of the material space. 
	This shift marks a significant departure from relying solely on chemical intuition, opening up new opportunities for materials discovery and design.~\cite{Pettifor_SSC_1984,Glawe_NJP_2016,oganov_JPCC_2020,Shires2021,Agrawal2016,Natrevstructure2019,Vonlilienfeld2020,Marzari2021}

	The key challenge for {\em in silico} material design is the prediction of
	thermodynamic stability of hypothetical compounds.
	This involves determining which element combinations will form stable phases under given conditions and understanding the types of structures they will adopt.~\cite{VandeWalle2002,Bartel2022}
	The traditional approach,  which combines {\em ab-initio} crystal structure prediction with convex hull construction, is accurate but computationally prohibitive for exploring binary, ternary, and multinary systems comprehensively.~\cite{Schoen1996,Goedecker2004,Oganov2006,Woodley2008,Ma2010,Oganov2011,Pickard2011,Wang2012,FloresLivas2020,Kolmogorov2021,Zurek2021}
	
	Data-driven high-throughput methods offer a more efficient, albeit less precise, alternative.~\cite{Curtarolo_Ceder_2003,Curtarolo2013,Genome_2013,Ghiringhelli_2015,Draxl_Scheffler_2018,Butler2018,Himanen2019,Schmidt2019} Techniques like element substitution on structural templates offer a way to explore extensive material spaces but depend heavily on the quality and representativity of the templates used.\cite{Fischer_2006,Hautier2011,Su2017,Sun2019,Merchant2023, Sun2019, Hautier2011, Wei2022, Kusaba2022, Santanu2023,cerqueira2024non} 
	
	In this paper, we propose a novel and efficient method,
	which combines the accuracy of ab-initio crystal structure prediction
	with the efficiency of data-driven techniques. 
	Our approach focuses on automatically constructing a minimal set of representative crystal structures (templates) to approximate convex hulls. The key innovation is that a carefully selected minimal set of templates can effectively capture the essential trends in chemical bonding and thermodynamic stability across the material space. This minimal set is generated through an iterative process that combines ranking vectors -assessing the thermodynamic stability of templates and their chemical behavior similarity - with clustering techniques to ensure diverse chemical coverage. This dual approach ensures that the final set of templates is both representative and efficient.
	
	Our preliminary tests show that our method dramatically reduces computational costs, achieving a reduction factor of approximately 25 compared to traditional EA searches. Despite this efficiency, it maintains high thermodynamic accuracy, with 80\% agreement with results from full EA searches. Remarkably, our method occasionally identifies more stable structures than traditional approaches, demonstrating its potential to outperform conventional methods in cases where EA searches are not exhaustive.

	While the focus of the current paper is on binary phase diagrams, our method  could be extended to ternary and even multinary systems.
	This scalability makes it a potentially transformative tool in materials discovery, enabling rapid and efficient exploration of complex chemical spaces that were previously computationally prohibitive. By bridging the gap between high-throughput efficiency and precise thermodynamic predictions, our method has the potential to accelerate the discovery of advanced materials across a wide range of applications, representing a significant advancement in ab-initio material design.

	The paper is organized as follows.
	Section \ref{sect:Idea} introduces the main concepts underlying our approach and gives formal definitions for each of them. Section \ref{sect:Algorithm} describes the practical implementation of the algorithm.  Section~\ref{sect:Results} describes the
	results of a preliminary application of the algorithm to 
	the study of binary convex hulls for the first 57 elements of the 
	periodic table (H-La) excluding hydrogen, including a partial validation on EA convex hulls. Finally, section~\ref{sect:Conclusions} contains the conclusions of the present work and discusses possible outlooks and perspectives.

	\section{Main Concepts and definitions:}\label{sect:Idea}  
	In this section we first give a short overview of our method and its goal. Then, we formally introduce the terms and concepts used throughout the paper. These will be used in the next section to describe the algorithm in detail. Some concept names are inspired by mathematical concepts. Despite these analogies, we make no claim of our description being mathematically rigorous, while we do rely on these analogies to ease the discussion.

	In short, our approach aims to obtain a small, meaningful set of crystal structures (templates) that can approximate the thermodynamic minimum for any pair of elements with reasonable accuracy. To achieve this, the template set must be carefully constructed to capture a wide range of chemical behaviors. 
	Our strategy to generate this set is an iterative procedure that relies on a limited number of {\em ab-initio} crystal structure prediction runs for representative element pairs and on the construction of generalized ranking vectors, based on the formation enthalpies computed on the templates. Ranking vectors permit to introduce a
	notion of distance between structures and chemical similarity between elements, guiding the efficient and meaningful selection of the template set.

	{\bf Materials Space:}  Within this paper, the term \textit{materials space} refers broadly to all potential bulk crystalline solids that can be formed  
	by selecting any two elements in any proportion from a list of
	$N_e$ elements(binary compounds) under a given external pressure. Each possible crystal structure represents a point in this
	generalized space, and it is possible to introduce notions of \textit{distance} between individual structures \cite{Valle_ActaCryst_2010_distance, Sadeghi_JCP_2013_metrics, Musil_ChemRev_2021_rep, cerqueira2024non}.
	
	This definition or material space implies that the chemical composition $x$ is unrestricted. Therefore, given the list of $N_{e}$ elements that define a materials space, the  number of possible distinct combinations is given by: $N_x \cdot N_e \cdot (N_e -1)$,
	where $N_x$ is equal to the number of compositions sampled.

\begin{figure}
		\centering
		\includegraphics[width=\linewidth]{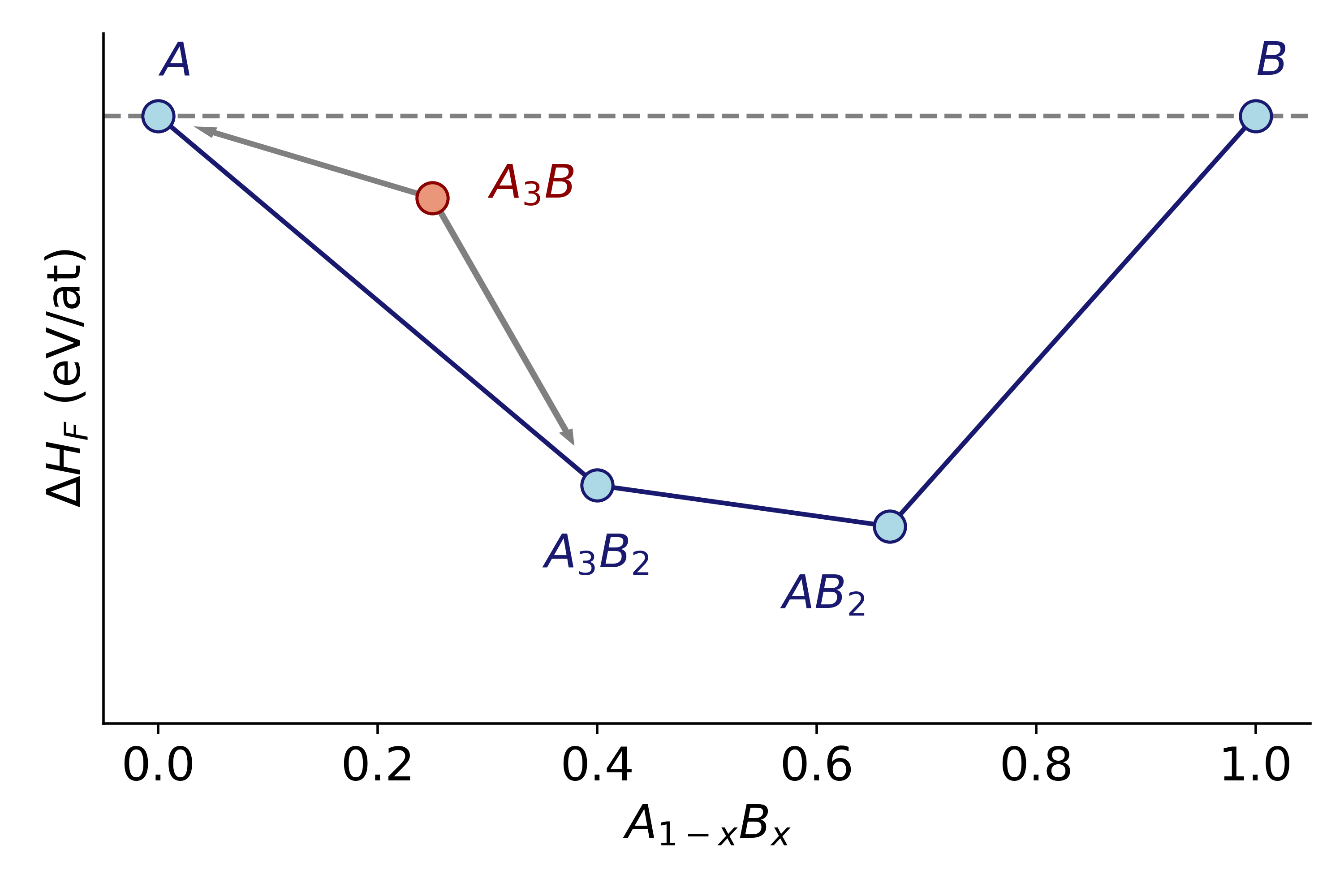}
		\caption{Convex hull construction for a binary \ch{A}, \ch{B} system.
			Blue and red symbols represent stable and metastable compositions, respectively. The grey arrows indicate possible decomposition paths of the \ch{A3B} phase into more stable compositions (\ch{A}+\ch{A3B2}).}
		\label{fig:convex_hull}
	\end{figure}

	{\bf Convex Hull:} 
	Given a set of compositions $N_x$ for an element pair, the  \textit{convex hull}
	construction is used to determine which ones correspond to stable structures.
	The construction of a binary convex hull for two elements $\ch{A}$ and $\ch{B}$ is illustrated in  Fig.~\ref{fig:convex_hull}.
	The convex hull is obtained plotting the formation enthalpies $\Delta H_F$ of the ground-state structures with compositions \ch{A_{1-x}B_x} as a function of $x$, and drawing the most convex curve that connects points at different $x$.
	
	Points lying on the convex hull (blue in Fig \ref{fig:convex_hull}) represent thermodynamically stable structures and compositions, while those lying above the hull (red) represent metastable
	phases that will decompose into different compositions ~\cite{Bartel2022}.
	The formation enthalpy $\Delta H_{F}$ for a structure with composition  \ch{A_{1-x}B_x}   is defined as:
	\begin{equation}
		\Delta H_{F} = H(\ch{A_{1-x}B_x}) - (1-x) H(\ch{A}) - x H(\ch{B}),
		\label{eq:Hf}
	\end{equation}
	where $H(\ch{A_{1-x}B_x})$, $H(\ch{A})$ and $H(\ch{B})$ are the enthalpies of the binary structure and of the ground-state structure of the two elements $A$ and $B$, respectively. The enthalpy for the pure elements was computed using the ground-state structures
	compiled in Ref. \cite{giannessi2024hex}.

	{\bf Templates:} 
	A template is a prototype crystal structure,  representative a specific chemical configuration. It is uniquely defined by its composition  \ch{A_{1-x}B_{x}}, its space group, and a set of Wyckoff positions. 
	Starting from each template, structures and enthalpies for different element pairs $P_{i}$ are obtained substituting different elements on the $A$ and $B$ lattice sites, 
	and relaxing the unit cell and internal parameters of the structure under constrained symmetry, up to a target pressure.
	Although it is in principle possible to assume arbitrarily complex structures as templates, practical considerations suggest to favor simpler structures that balance representativeness with computational cost. 
	
	{\bf Basis set:} 
	Within the \textit{material space}, we  define
	a generalized \textit{basis set} as the collection of all structures that correspond
	to minima of the free energy, i.e. all thermodynamically stable structures.
	The convex hulls computed using this \textit{full basis set} would, by definition, be exact.
	However, obtaining this basis set would require predicting ground-state
	structures for all possible element pair combinations, which is computationally unfeasible.
	This paper assumes that this large, unknown basis set, 
	can be approximated by a much smaller {\em minimal basis set} of prototype structures (templates). These templates
	collectively capture the chemical behavior and structural diversity across all relevant element pairs within a specified regime of pressure and composition.
	Careful construction of this minimal basis set 
	allows for very efficient exploration of material space, without considerable loss of accuracy.

\begin{figure}
		\centering
		\includegraphics[width=\linewidth]{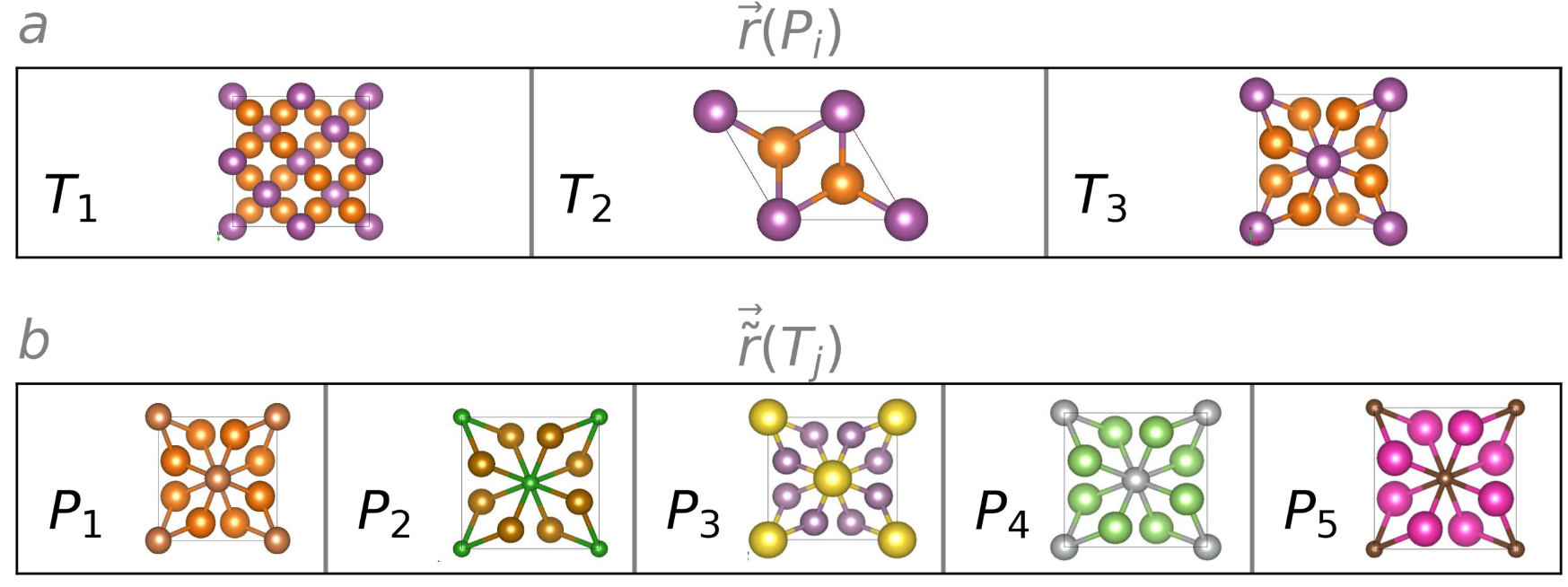}
		\caption{Definition of the two types of ranking vectors employed in this work.
			Panel $a$): Pair ranking vectors.
			For each element pair $P_i={A_i,B_i}$ the elements of $\vec{r}(P_i)$ represent the templates $T_{j}$ ordered in terms of ascending formation enthalpy.
			For example, in the ranking vector shown in the figure $\Delta H_F\left(T_1\left(P_i\right)\right) \textless \Delta H_F\left(T_2\left(P_i\right)\right) \textless \Delta H_F\left(T_3\left(P_i\right)\right)$ etc. 
			Panel $b$): 
			Template ranking vector $\vec{\tilde{r}}(T_{j})$.
			Given a single template $T_{j}$, the 
			elements of $\vec{\tilde{r}}(T_{j})$ contain different element pairs $P_i$, ranked in order of ascending formation enthalpy. 
			For example, in the ranking vector shown in the figure,
			the pair $P_1$ has the lowest formation enthalpy on $T_{j}$, $P_2$
			the second lowest, and so on.}
		\label{fig:ranking}
	\end{figure}

	{\bf Ranking Vectors:}
	A {\em ranking vector} is a vector used to rank crystal structures with the same composition $x$ on the basis of their formation enthalpies $\Delta H_{F}$. 
	Introducing ranking vector permits to define a notion of distance in the material space.  In this paper we define two types of ranking vectors:
	
	\begin{itemize}
		\item A {\em Pair ranking  vector} $\vec{r}(P_i)$ (Fig.~\ref{fig:ranking} $a$)
		is constructed for a fixed element pair $P_i$. Each component of the vector 
		represents a single structural template. The templates are
		ranked in ascending order of formation enthalpy. In other words, this vector answers the question: for this given pair of elements $P_{i}$, how are the templates $T_{j}$ ranked, from lowest to highest formation enthalpy? 
		Pair ranking vectors are used to assess the  {\em chemical similarity} between element pairs. We assume that if two or more element pairs have similar chemical behaviour,
		they will prefer or avoid the same type of structures, resulting in similar ranking vectors. Conversely, chemically different elements will prefer 
		different types of structures.
		\item 
		A {\em Template ranking vector} $\vec{\tilde{r}}(T_{j})$ 
		(Fig.~\ref{fig:ranking} $b$) is constructed for a single structural template $T_{j}$. The components  are element pairs, ranked in the order of their
		formation enthalpy on that template.
		This vector answers the question: given the template $T_{j}$, how are pairs of elements $P_{i}$ ranked, from lowest to highest formation enthalpy?
		
		Template ranking vectors are useful to assess redundancy among templates. Similar ranking vectors indicate that two templates tend to describe effectively the same element combinations and can be considered redundant when aiming for efficient representation of the material space.
	\end{itemize}

	{\bf Levenshtein Distance:}
	The Levenshtein distance is a metric used to measure the distance between two strings \cite{Levenshtein1965}, by counting the minimum number of changes required to turn one string into another.
	In the context of this paper, the strings represent ranking vectors of templates or element pairs. Hence,  the Levenshtein distance is used to
	quantify the distance between element pairs (chemical similarity) or  between templates (redundancy). In fact, a short distance between pair ranking vectors indicate
	that two element pairs have similar bonding behavior, while a short distance
	between template ranking vectors indicate that two templates  likely 
	represent similar structural environments.

	For two strings of equal length $N$ the weighted Levenshtein distance is defined as:
	\begin{equation}
		d = \sum_{i=0}^{N-1}\omega_ix_i.
		\label{eq:dlev}
	\end{equation}
	$x_i$ is an integer, which is zero if the element entry $i$ is identical in the two strings, or one if it is different. $\omega_i$ is a weighting factor, defined as: 
	\begin{equation}
		\omega_i = 1 - \dfrac{i}{N} ~~ i \in [0,N-1],
	\end{equation}
	which reflects the fact that elements at the beginning of the vector (representing templates or pairs with lower formation enthalpies) are more significant than those towards the end.
	
	To compare distances of ranking vectors of different lengths (due to varying numbers of templates or element pairs sampled), we use a normalized Levenshtein distance $\tilde{d}$, defined as: 
	\begin{equation}
		\tilde{d} = \dfrac{1}{d^{max}}\sum_{i=0}^{N-1}\omega_i x_i,
		\label{eq:maxlev1}
	\end{equation}
	where
	\begin{equation}
		d^{max} = \sum_{i=0}^{N-1}\omega_i
		\label{eq:maxlev2}
	\end{equation}
	is a normalization factor ensuring that $\tilde{d}$ ranges between 0 and 1.
	
	{\bf Clustering and Chemical Similarity:}
	To enhance the efficiency of the template generation procedure, we employ agglomerative clustering to partition the material space into regions (clusters) of chemically-similar element pairs. The assumption is that templates generated by element pairs from different regions  will likely differ significantly. By learning this partitioning of the material space, we can reduce the number of evolutionary algorithm (EA) searches needed to generate templates by selecting element pairs from different regions.
	The concept of a minimal basis set 
	of structural templates is closely linked to the idea that the material space can be divided into a limited number of regions, each characterized by a specific type of chemical interaction. Thus,  a \textit{minimal basis set} must include templates representative of element pairs from different clusters, which are chemically distinct.
	
	{\bf EA calculations/runs}
	Throughout the text, we use the term EA (Evolutionary Algorithm) to refer to crystal structure prediction calculations performed using the \verb|USPEX| code \cite{Oganov2006, Oganov2011, Oganov2013_2}, which implements genetic algorithms for structure prediction, and \verb|Quantum ESPRESSO| \cite{Giannozzi2009, Giannozzi2017} to perform structural relaxations and compute the total enthalpy. Computational details are given in Sect. I of the Supplementary Material \cite{suppmat}.
	
	%%%%%% IMPLEMENTATION %%%%%%%%%%%%%%%
	
	\section{Description of the Algorithm}\label{sect:Algorithm}
	
\begin{figure*}[!htbp]
		\centering
		\includegraphics[width = 0.9\linewidth, trim={2cm 0cm 2cm 0cm}]{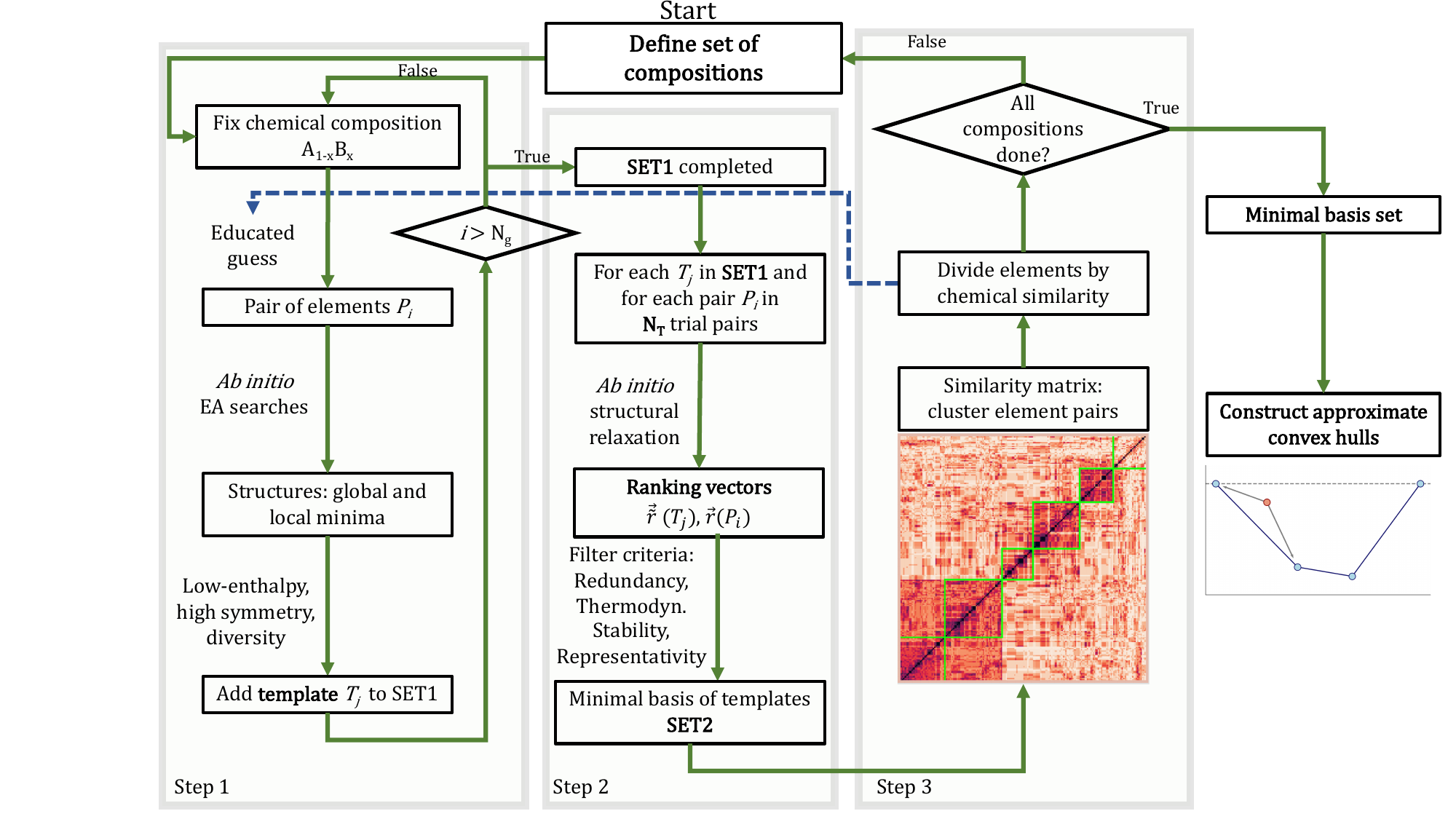}
		\caption{Flowchart illustrating the template generation procedure.}
		%The template generation can be roughly divided into two main phases, an \textbf{Initialization at First Composition} and then the \textbf{Extension to Other Compositions}, each refined by a \textbf{Statistical Analysis} phase.}
	\label{fig:flowchart}
\end{figure*}

	We start by giving a general overview of the algorithm to automatically generate a minimal basis set of templates, followed by a detailed description of each step. A general flowchart of the algorithm is provided in Fig.~\ref{fig:flowchart}. 
	
	The algorithm samples sequentially several user-defined compositions $x$ sequentially to construct the convex hull. 
	%STEP 1:
	For each composition, in the {\bf first step} a large set of candidate templates,
	referred to as SET1, is generated first, by performing
	sequential EA searches for a set of $N_g$ randomly-chosen
	{\em generating element pairs}. Templates are selected from among the lowest-enthalpy structures 
	identified by these searches, taking into account symmetry,
	diversity, and thermodynamic stability.
	
	%STEP 2:
	After generating SET1,  in the {\bf second step} a large number $N_T$ of additional element pairs -- {\em trial element pairs} -- are relaxed on all templates in SET1.
	The results of the relaxations are used to
	construct pair and template ranking vectors, which are used to evaluate the performance of each template in terms of {\em redundancy}, {\em thermodynamic stability} and {\em representativity}. 
	The best-performing templates, which most 
	effectively represent the structural and energetic characteristics of the materials space,
	are selected to form the {\em minimal basis set} (SET2) for the composition.
	
	%STEP 3
	
	Once the minimal basis set of templates has been constructed, in the {\bf third step} the element pairs are divided into a small number of $N_c$ clusters based on their chemical similarity, as evaluated through the pair ranking vectors.
	Clustering primarily guides the selection of optimal element pairs during the template generation phase for subsequent compositions.

\subsection*{Step 1: Generation of the Initial Set of Templates:}
To ensure an efficient generation of a representative initial template set, the $N_g$ generating element pairs need to be chosen from sufficiently diverse regions of the materials space. For all compositions beyond the first,  the $N_g$ pairs 
are chosen
from the $N_c$ different clusters of the pair similarity matrix obtained at previous compositions. 
For the first composition, where previous clustering information is unavailable, we employ a bootstrapping procedure, based on {\em reference vectors},
which guide the selection of element pairs spanning a broad range of chemical tendencies and structural motifs. The bootstrapping ensures the algorithm starts with a robust and diverse set of initial templates.

\subsubsection{First Composition:}
The initial set of of templates for the first composition is
generated using  {\em reference vectors} to label chemically-different regions of the material space.
The list is updated as more element pairs are sampled during the template generation phase.

The algorithm is initialized by selecting a random pair of elements  $P_1=(A,B)$ from the pool of $N_e$ elements and performing a fixed-composition EA search for this pair. 
The two lowest-enthalpy, high-symmetry structures identified by the search are chosen as initial candidate templates $T1$ and $T2$, and the pair ranking vector $\vec{r}(P_1)=[T1,T2]$ is chosen as the first reference vector $\vec{R}_1$.

Next, a second pair of elements $P_2=(C,D)$ is randomly selected. The elements $C$ and $D$ are substituted into the initial templates $T1$ and $T2$ to obtain a new ranking vector $\vec{r}(P_2)$. This new ranking vector is then compared with the initial reference vector $\vec{R}_1$. If  $\vec{r}(P_2)\neq \vec{R}_1$, $P_2$ is assumed to be chemically diffent from $P_1$ and a fixed-composition EA search is performed for $P_2$, to identify potentially a new template $T_3$. For this, high-symmetry structures within a fixed enthalpy range above the ground state are screened (see SM for details). 
If a suitable new structure is found, it is added to the template set as $T3$, and  $\vec{R}_2=\vec{r}(P_2)$ is added to the list of reference vectors. All reference vectors are updated to include $T3$.

However, if $\vec{r}(P_2) = \vec{r}(P_1)$, or if no suitable new template is found, another random pair of elements $P_3=(E,F)$  is selected, and the process is repeated. 
The process repeats iteratively, selecting pairs and updating the template set and reference vectors until no new templates are found or the template set is complete.

\subsubsection{Other Compositions}
For all compositions beyond the first, templates generation leverages the clustering results from previous compositions, to ensure that the EA searches yield templates from chemically distinct regions of the materials space.
The $N_g$ generating element pairs are selected by drawing $N_{g_i}$ pairs from each of the $N_c$ clusters identified during the clustering phase, along with $N_{g_+}$  additional pairs from the positive enthalpy cluster.
This method is more efficient than the bootstrapping procedure, requiring fewer searches to reach the same number of templates.

With this initial set of candidate templates (SET1), we ensure diverse chemical and structural representation, laying the foundation for further refinement based on element pair relaxations. This diversity is critical for accurately representing the full breadth of material behaviors within the targeted compositions, and it lays the foundation for capturing the full thermodynamic behavior of the material space, ultimately enhancing the accuracy of the final convex hull predictions.

\subsection*{Step 2: Reduction to a Minimal Basis Set:}
In this phase, the initial set of candidate templates  (SET1) is reduced to a minimal basis set (SET2) through  statistical analysis performed on a large number $N_T$ of {\em trial element pairs} ($N_T \leq N_e$).

A  database of structures, element pairs and enthalpies is created substituting the $N_T$
trial element pairs into all templates in SET1.
The corresponding structures are then relaxed using DFT to a common pressure using the same convergence criteria (See Supp. Mat. section I. B. for further details), and their formation enthalpies are computed. Template and pair ranking vectors, $\vec{r}(T_{j})$, $\vec{r}(P_i)$, are constructed as outlined in section \ref{sect:Idea}.

The trial element pair relaxations provide critical data for constructing ranking vectors that quantify the performance of each template. These vectors serve as the basis for evaluating templates in terms of redundancy, thermodynamic stability, and representativity, setting the stage for reducing the template set 
from a broad set (SET1) to a minimal basis set (SET2).

In particular, we used  the following \textbf{filter criteria} based on  ranking vectors:
\begin{itemize}
	\item {\em Redundancy}:
	Templates that behave similarly in terms of bonding and thermodynamic stability are removed, as they add little new information, and hence can be considered redundant.
	Redundancy is evaluated by computing the weighted Levenshtein distance between the ranking vector
	$\vec{\Tilde{r}}(T_{j})$ and those of other templates $T_{i}$ in the basis set. 
	A short distance between two templates implies that they tend to describe the same sets of elements similarly. 
	\item {\em Thermodynamic stability}: Templates that consistently yield stable structures across different element combinations are prioritized.
	This is assessed by counting how often
	$T_{j}$ exhibits a negative formation enthalpy for the $N_T$ trial element pairs.
	\item {\em Representativity}: We wish to retain templates that are representative, i.e.
	which  capture a unique bonding environment or phase behavior that is representative of a subset of the material space.
	This is measured by counting how often $T_{j}$ appears as the first entry in a pair ranking vector, i.e. how often it occurs as a ground-state structure.
\end{itemize}

Based on a combination of these three criteria, templates in SET1
are either retained in SET2, or discarded.
This ensure that the retained templates in SET2 not only provide a diverse and representative structural set but also maintain thermodynamic accuracy across the sampled element pairs.
Further details on the reduction procedure, including computed results on redundancy, thermodynamic stability, and representativity for the test case described in the second part of this manuscript, can be found in the Supplemental Material section II.B.

By applying the filtering criteria based on redundancy, thermodynamic stability, and representativity, we reduce the initial set of templates (SET1) to a more manageable and efficient minimal basis set (SET2). This step ensures that only the most informative templates are retained, maintaining both accuracy and computational efficiency.

\subsection*{Step 3: Clustering:}
Once the templates in SET1 have been reduced to SET2, we
construct the pair ranking vectors $\vec{r}(P_i)$ for all $N_T$ element pairs, using only these templates. We then construct the similarity matrix (See Ref. \cite{Gower_Biometrics_1971_similarity}) using normalized, weighted Levenshtein distances ($\tilde{d}$) between these pair ranking vectors $\vec{r}(P_i)$. 
The resulting matrix is transformed into a dendrogram, which divides the element pairs into a smaller number of $N_c$ clusters. The heigth of the dendrogram, and hence the number $N_c$ of clusters, can be adjusted based on the desired level of detail in understanding specific regions within the chemical space (See Sect. II.C. for further details on the clustering method).
In our algorighm, clustering element pairs based on chemical similarity 
serves to
ensure that the templates in 
the final set (SET2) capture a broad range of chemical behaviors.
We found that a relatively small number of clusters ($N_c=6$), i.e. about one cluster per template per composition, plus one for outliers, is an optimal choice to 
enhance the efficiency of the method without sacrificing accuracy.

\section{Preliminary application:}\label{sect:Results}
In this section, we describe a first application of the method outlined in the previous two sections. First, we will list a few key parameters defining the material space,
i.e. the choice of elements, the  set of compositions, number of pairs to be generated, selection criteria and so on. Then, we will present the set of templates obtained using these parameters. Finally, we will  validate our results by comparing how the minimal basis set of templates performs against full ab initio crystal structure prediction EA searches in describing the convex hull for a large set of test element pairs.

\subsection{Choice of parameters:}
We considered the materials space consisting of the first 57 elements of
the periodic table (excluding hydrogen), spanning from helium to lanthanum, 
at a pressure of 200 GPa. We restricted ourselves to four possible compositions (\ch{AB1}, \ch{AB2}, \ch{AB3}, \ch{AB6}), which correspond to seven points on the convex hulls (due to A/B permutation); \ch{AB2} was selected as the initial composition, and then 
we proceeded with other compositions in the order \ch{AB3}, \ch{AB1} and \ch{AB6}. 

The parameters of the template generation procedure for each composition are provided in Table \ref{tab:data_template_reduction}. For each EA search at a given composition, exctept from the first,  we aimed to select one template from the identified structures that satisfied the following criteria:
i) the structure lies up to 400 $meV /at$ above the ground-state; ii) it has a space group higher than 75 (tetragonal, trigonal, hexagonal or cubic Bravais lattice) and iii) the 
same template (in terms of space group and Wyckoff positions) is not already included in
the set. If no structure met these criteria, a new element pair was selected. All EA searches employed unit cells containing up to 2 f.u., except for the $1:1$ composition, where cells up to  4 f.u. were allowed.

\begin{table}[]
	\centering
	\begin{tabular}{|c|c|c|c|c|}
		\hline
		composition & $N_g$ & $N_T$ & $N$(set$_1$) & $N$(set$_2$) \\ 
		\hline
		\ch{AB2}* & 80 & 330 & 17 & 5 \\
		\ch{AB1} & 33 & 268 & 7 & 4 \\
		\ch{AB3} & 26 & 207 & 9 & 5 \\
		\ch{AB6} & 28 & 138 & 8 & 4 \\
		\hline
	\end{tabular}
	\caption{Details of the template generation procedure for the different compositions
		sampled in this work.
		Number of generating ($N_g$) and trial ($N_{T}$)
		element pairs used in the template generation and reduction
		procedure. $N$(SET1) and $N$(SET2) indicate the number of templates 
		in the initial set and in the final, minimal basis set, respectively.
		The asterisk indicates that \ch{AB2} is the starting composition, for
		which SET1 was generated using the procedure described in Sect. II.
	}
	\label{tab:data_template_reduction}
\end{table}

Once SET1 was constructed for one composition, we relaxed $N_T$ different element pairs on the templates of SET1 and ranked them by redundancy, thermodynamic stability, and representativity. In addition to  these criteria, we prioritized structures with smaller unit cells and higher symmetry. Using these filter criteria, we reduced the basis set from SET1 to SET2. For further details on the criteria used to reduce the number of templates and its impact on accuracy, see the Supplemental Material Sections II.A. and III, respectively.

\subsection{Final set of templates:}
\label{sect:Templates}
We finally obtained the minimal basis set of templates containing only 18 structures across the 4 chosen compositions (\ch{AB}, \ch{AB2}, \ch{AB3}, \ch{AB6}). As we will demonstrate in Sect. \ref{sect:Validation}, this very small set of structures is able to describe the thermodynamics of most of the elements with remarkable accuracy.

\begin{figure*}
	\centering
	\includegraphics[width=\linewidth, trim={0cm 1.5cm 0cm 0cm}]{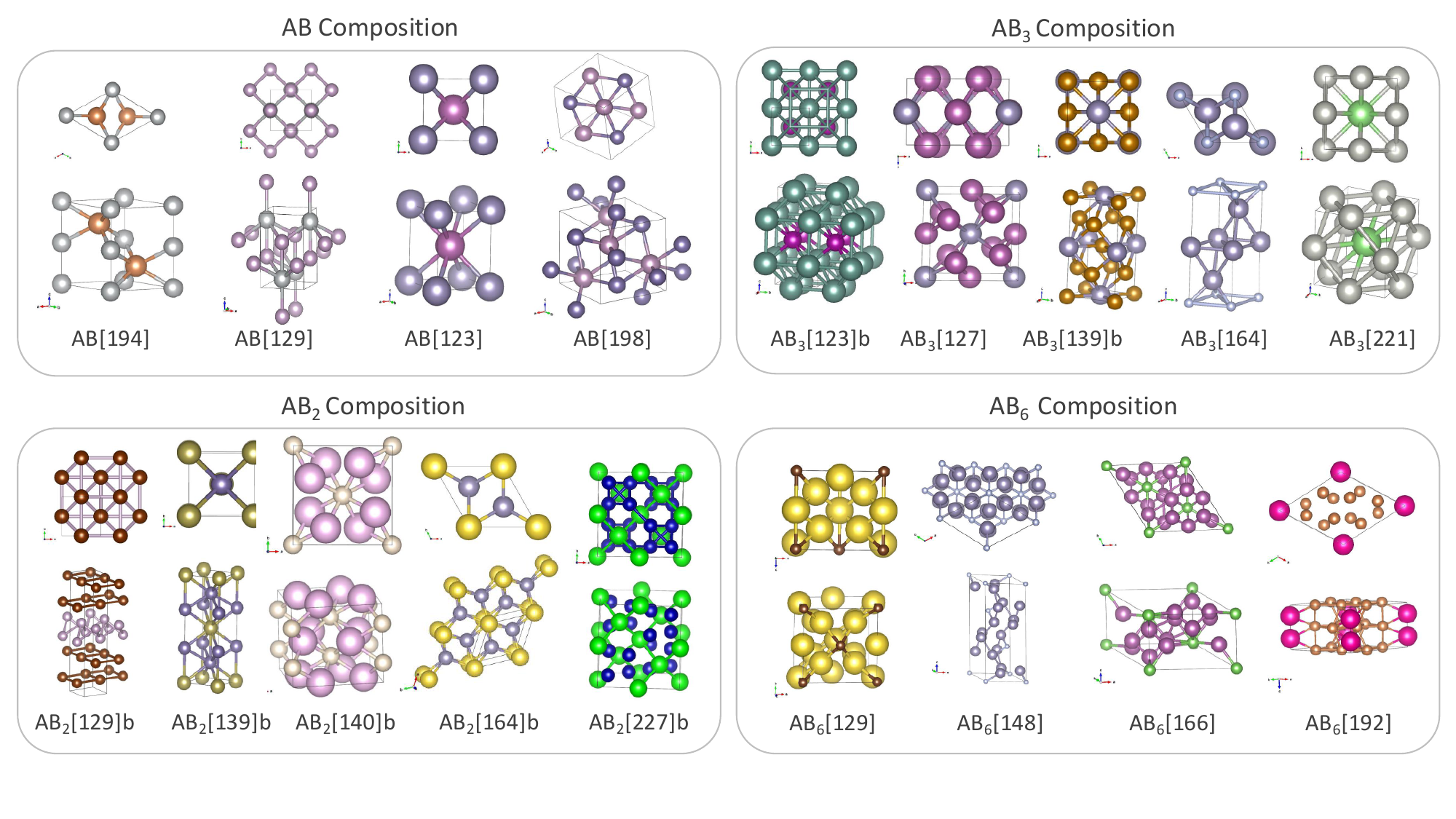}
	\caption{Top and side views of the unit cells of all templates forming the final set at 200 GPa. The corresponding structural parameters are reported in in table \ref{tab:stucturalinformation}.
		Figures were generated with the VESTA software.}
	\label{fig:crystalStructures}
\end{figure*}

Top and side views of the unit cell of all templates are shown in Fig. \ref{fig:crystalStructures}, along with their space group number. Table \ref{tab:stucturalinformation} summarizes the relevant structural information. The first column provides the unique ID of each structure, indicating the composition, space group number (in square brackets), and a lowercase letter to distinguish structures with the same space group. The second column lists the element pair that originally generated the structure through EA searches, as described in the previous section. The last column contains the
type and multiplicity of Wyckoff positions. 

\begin{table}[b]
	\begin{tabular}{| c | c | c | l |}
		\hline
		ID & Elements & \begin{tabular}{ c } space \\ group \\ \end{tabular} & Wyckoff Positions \\ \hline
		$AB[194]$ & SbNi & 194  & \textbf{A}:\textit{2c} \textbf{B}:\textit{2a} \\
		$AB[129]$ & PNi & 129  & \textbf{A}:\textit{2c}  \textbf{B}:\textit{2c}  \\
		$AB[123]$ & SnSc & 129  & \textbf{A}:\textit{1a} \textbf{B}:\textit{1d} \\
		$AB[198]$ & GeMo & 198 & \textbf{A}:\textit{4a} \textbf{B}:\textit{4a}  \\
		\hline
		$AB_2[129]b$ & PBr & 129 & \textbf{A}:\textit{2c}  \textbf{B1}:\textit{2c} \textbf{B2}:\textit{2c} \\
		$AB_2[139]b$ & TeGe & 139  & \textbf{A}:\textit{2a}  \textbf{B}:\textit{4e} \\
		$AB_2[140]b$ & HeKr & 140  & \textbf{A}:\textit{4a}  \textbf{B}:\textit{8h} \\
		$AB_2[164]b$ & NaSn & 164  & \textbf{A}:\textit{1a}  \textbf{B}:\textit{2d} \\
		$AB_2[227]b$ & ZrCo & 227  & \textbf{A}:\textit{8b}  \textbf{B}:\textit{16c}\\
		\hline
		$AB_3[123]b$ & MnY & 123 & \textbf{A}:\textit{1d}  \textbf{B1}:\textit{1c} \textbf{B2}: \textit{2g}  \\
		$AB_3[127]$ & SnSc & 127  & \textbf{A}:\textit{2b}  \textbf{B1}:\textit{2c} \textbf{B2}:\textit{4g} \\
		$AB_3[139]b$ & SnFe & 139  & \textbf{A}:\textit{2b}  \textbf{B1}:\textit{2a} \textbf{B2}:\textit{4d} \\
		$AB_3[164]$ & NSn & 164  & \textbf{A}:\textit{1a}  \textbf{B1}:\textit{1b} \textbf{B2}:\textit{2d} \\
		$AB_3[221]$ & LiPd & 221 & \textbf{A}:\textit{1b}  \textbf{B}:\textit{3d}\\
		\hline
		$AB_6[129]$ & CNa & 129 & \textbf{A}:\textit{2a}  \textbf{B1}:\textit{2c}  \textbf{B2}:\textit{2b} \textbf{B3}:\textit{8j} \\
		$AB_6[148]$ & NSn & 148 & \textbf{A}:\textit{3a}  \textbf{B}:\textit{18f} \\ 
		$AB_6[166]$ & AsSc & 166 & \textbf{A}:\textit{3a}  \textbf{B}:\textit{18h} \\ 
		$AB_6[192]$ & RbSb & 192 & \textbf{A}:\textit{2a}  \textbf{B}:\textit{12l} \\
		\hline
	\end{tabular}
	\caption{Structural Parameters of the final template set for 200 GPa: Unique ID (composition [Space group]), space group, type and multiplicity of Wyckoff positions. Top and side views of the
		corresponding unit cells are 
		shown in  Fig.~\ref{fig:crystalStructures}.}
	\label{tab:stucturalinformation}
\end{table}

\subsection{Validation:}\label{sect:Validation}
With the minimal basis set (SET2) constructed, we now validate our method by comparing the results to those obtained from full ab-initio EA searches. This validation will demonstrate the efficiency and accuracy of our approach across a wide range of element pairs and compositions.

We constructed EA {\em validation hulls}
for 100 element pairs. For each pair, we performed fixed composition
EA searches  for the A$_6$B, A$_3$B, A$_2$B, AB, AB$_2$, AB$_3$ and AB$_6$ compositions, using reasonably tight convergence criteria. Detailed information on these criteria can be found in the Supplemental Material (SM Sect. III, Tab. III \cite{suppmat}). 

\begin{figure*}
	\centering
	\includegraphics[width=0.95\linewidth]{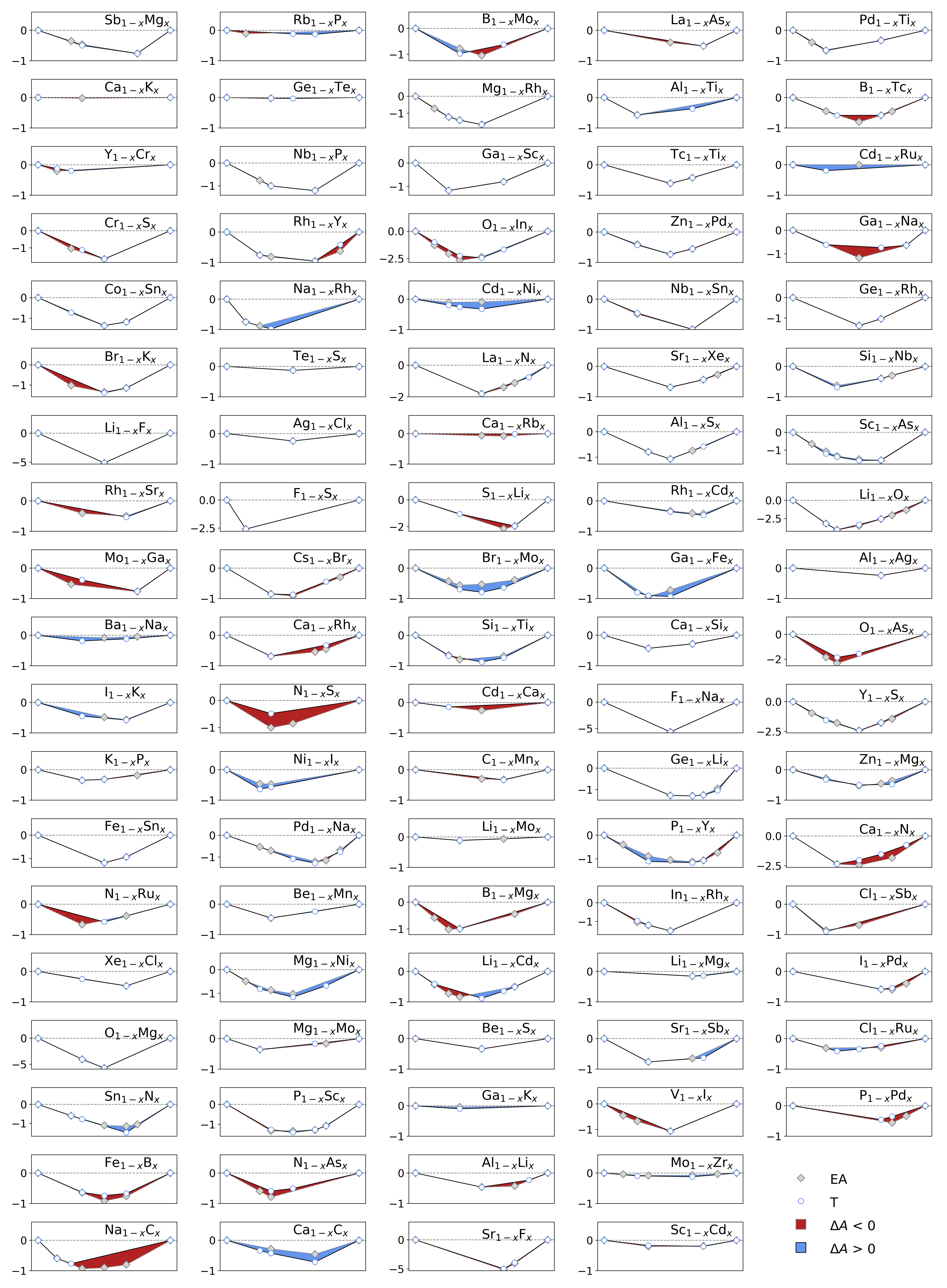}
	\caption{Comparison between the convex hulls predicted with our template  method (T) and with evolutionary algorithms (EA) for a 
		validation set of 100 element pairs. 
		We only show pairs for which the EA predicts a hull with a finite (negative) depth.
		The \textit{x}- and \textit{y}-axes
		of each graph represent the composition $x$ and the formation enthalpy $\Delta H_F$ in  eV/at respectively -- labels are omitted for better visibility. Blue and red-shaded areas indicate regions where  the T (EA) hull is \textit{below} the EA (T) hull respectively.}
	\label{fig:validation}
\end{figure*}

Fig.~\ref{fig:validation} presents a comparison between the results obtained using the evolutionary algorithm (EA, grey diamonds) and those from the template method (T, blue circles). The red (blue) shaded areas highlight instances where the EA hull is below (above) the T hull, respectively.

Overall, there is a notable agreement between the two methods in both the depth and shape of the convex hulls. Specifically, 80\% of the stable/unstable compositions are predicted correctly over all the convex hulls, and the depth of the T hull (i.e. the enthalpy of the lowest point) is not higher than 50 meV above the EA hull.
In some cases, the T curve is even lower than the EA curve, i.e. template structures are thermodynamically more stable than those identified through EA searches.
This suggests that the template method might surpass EA searches when exploring broad chemical spaces, particularly if the EA search is not exhaustive.
It is worth to stress that while 120 structural relaxations were needed to determine each point on the EA validation hull, only 4-5 relaxations were required per composition for T hulls, reducing the computational time by a factor 25. 
In synthesis, our validation demonstrates that the template method not only reduces computational time by up to 1/25 compared to traditional EA searches but also maintains thermodynamic accuracy, making it a powerful tool for large-scale materials exploration.

\section{Conclusions:}\label{sect:Conclusions}
In this paper, we introduced a novel method for efficiently exploring binary phase diagrams through the automatic construction of a minimal basis set of representative templates to approximate convex hulls. This method requires no prior knowledge of the material space; instead, a limited number of crystal structure prediction runs are performed for selected element pairs, and these templates are refined through a data-driven process.

A core element of our approach is the concept of the existence of a minimal basis set of templates. The underlying assumption is that the material space can be partitioned into regions with similar chemical tendencies and bonding properties, and that a small set of representative templates can effectively capture these trends. This minimal basis set can be viewed as analogous to a basis in a generalized Hilbert space. By employing statistical tools such as ranking vectors, Levenshtein distance, and clustering analysis, we quantitatively assess template similarity and ensure that the final set of templates provides comprehensive coverage of chemically diverse regions.

Preliminary results validate our initial assumption and demonstrate the effectiveness and efficiency of this approach. Tests on the first 57 elements of the periodic table (excluding hydrogen) at 200 GPa show that our template-based method accurately reproduces convex hulls predicted by computationally intensive Evolutionary Algorithm (EA) searches for approximately 80\% of the sampled element pairs. This is achieved with a computational cost reduction of about 25 times, underscoring the potential for significant resource savings while maintaining high accuracy.

This approach provides a crucial solution to the growing demand for more efficient methods in materials discovery. Further systematic testing is required to evaluate the robustness and versatility of the method across a broader range of binary phase diagrams, but its extension to ternary and multinary systems—where computational efficiency is even more critical—could dramatically transform the pace of discovery. By providing a scalable and precise tool for navigating the complexities of multi-component systems, our method offers an urgently needed breakthrough in high-throughput materials discovery, opening the door to generating unbiased material maps and extending traditional classification systems like Mendeleev’s periodic table or Pettifor material maps.~\cite{Pettifor_SSC_1984,Glawe_NJP_2016,Rastrepo_PNAS_2022,oganov_JPCC_2020,Shires2021}

Ultimately, our method has the potential to revolutionize in silico material design and unlock new pathways for discovering advanced quantum materials, driving forward technological innovations at a scale and speed previously considered unattainable.

\section{Acknowledgments}
The authors acknowledge support by the state of Baden-Württemberg through bwHPC
and the German Research Foundation (DFG) through grant no INST 40/575-1 FUGG (JUSTUS 2 cluster).
L.B. and. S.D.C. acknowledge support from
Fondo Ateneo Sapienza 2019-22, and funding from the European Union - NextGenerationEU under the Italian Ministry
of University and Research (MUR), “Network 4 Energy Sustainable Transition - NEST” project (MIUR project code PE000021, Concession Degree No. 1561 of October 11, 2022) - CUP C93C22005230007.
%\printbibliography

\bibliographystyle{naturemag}
%\bibliography{main}

\end{document}